\documentstyle[12pt,worldsci]{article}
\begin{document}
{\vspace*{-1cm}
\flushleft
{\tt To be published in {\em Electronic Surface and
Interface States on Metallic Systems}, eds.: E. Bertel und M. Donath
(World Scientific, Singapore, 1995).}
       \vspace{1cm}\\}
\title{
Frustrated H-induced Instability of Mo(110)}
\author{P. Ruggerone, B. Kohler, S. Wilke, and M. Scheffler\\
{\em Fritz-Haber-Institut der Max-Planck-Gesellschaft, Faradayweg 4-6,
D-14195 Berlin-Dahlem, Germany}}
\maketitle
\begin{abstract}
The large interest in the properties of transition metal surfaces has
been recently fostered by inelastic helium atom scattering
experiments carried out by Hulpke and L\"{u}decke. Sharp and giant
anomalies in the surface phonon dispersion curves along $\overline {\Gamma {\rm
H}}$
and $\overline {\Gamma {\rm S}}$ have been detected on W(110) and
Mo(110) at a coverage of one monolayer of hydrogen. At the same critical
wave vectors a smaller second indentation is present in the experimental
phonon branches. Recently we have proposed a possible
interpretation which is able to explain
this and other experiments. In this paper we
discuss results of our recent ab initio calculations of the atomic and
electronic
properties of
the clean and H-covered Mo(110) surface in more details. For the full
monolayer
coverage
the calculated Fermi-surface contours are characterized by strong
nesting features which originate the observed anomalies.
\end{abstract}
\section{I\lowercase{ntroduction}}
The original goal of the inelastic helium atom scattering (HAS) investigation
on H/W(110) performed by Hulpke and L\"{u}decke~\cite{HUL92,HUL93,LUE94} was
to find a lattice dynamical fingerprint of the so called
{\it top-layer-shift reconstruction}~\cite{EST86}. This structural model
was proposed by Estrup and coworkers~\cite{EST86} to explain their low
energy electron diffraction (LEED) data with hydrogen adsorbed on the
surface:
the presence of hydrogen
induces a rigid shift along the [${\overline 1}10$] direction
of the top layer atoms with respect to the substrate.
However, the size of this suggested displacement is still under discussion
and no evidence for such a rearrangement has been reported for
H/Mo(110)~\cite{ALT87}. The HAS data did not offer insight into the
mechanism of the top-layer-shift reconstruction, but were quite amazing.
On W(110) and Mo(110) the HAS data for the clean phase do not exhibit any
peculiarities, whereas
for a full monolayer of hydrogen the surface phonon dispersion curves are
characterized by two anomalous indentations which both occur at the same
critical wave
vector ${\bf Q}_{\rm c1}^{\rm exp}$ along the [$001$] direction
($\overline {\Gamma {\rm H}}$). ${\bf Q}_{\rm c1}^{\rm exp}$
has a length of 0.95~\AA$^{-1}$ for W and  0.90~\AA$^{-1}$ for
Mo. The dips have been also detected along directions deviating
from $\overline {\Gamma {\rm H}}$. In particular, along
the [${\overline 1}12$] direction ($\overline {\Gamma {\rm S}}$)
the indentations occur at the $\overline {{\rm S}}$ point for Mo and very close
to it for W,
having the critical wave vector ${\bf Q}_{\rm c2}^{\rm exp}$ length of 1.225
and
1.218~\AA$^{-1}$ for W and Mo respectively. Since the experimental curves
are available only for $\overline {\Gamma {\rm H}}$ and
$\overline {\Gamma {\rm S}}$ (see Ref.~3), we focus our
attention on these two cases characterized by strong and well defined
anomalies in the phonon branches. For both systems two different
dips in the surface dispersion curves were observed at the
critical wave vectors: a smaller softening and a very sharp and localized
indentation from $\hbar \omega \approx 15$~meV to $\hbar \omega \approx 2$~meV.
Similarly deep and sharp anomalies in the phonon branches
have been detected
only in neutron scattering experiments on quasi-one-dimensional
conductors, like KCP [K$_2$Pt(CN)$_4$Br$_3$]~\cite{COM75}, but never
on surfaces. A pronounced but less sharp damping of
longitudinal surface phonons was detected on the (100) surfaces
of W~\cite{ERN,SCH89} and Mo~\cite{HUL89} and was interpreted as a soft
phonon due to a Kohn anomaly involving electronic surface states at
the Fermi level~\cite{SMI90,CHU92}. In the case of W(100) the Fermi
surface nesting is particularly efficient and drives a structural
rearrangement of the atomic pattern leading to a c($2 \times 2$)
reconstruction at temperatures below $T_{\rm c}\approx 250$~K (for a
review see Ref.~\citenum{JUP94}). As already mentioned, the phonon dispersion
curves for the clean (110) surfaces do not present any unusual features;
the anomalies appear only when one monolayer
of hydrogen is adsorbed. However, it seems that hydrogen vibrations do
not play a crucial role in the appearance of the indentations, because the
HAS spectra do not change when deuterium is adsorbed instead
of hydrogen~\cite{HUL93,LUE94}. Moreover, a possible link between the
anomalies and the top-layer-shift reconstruction should be ruled out
because the dips were also observed on H/Mo(110) which does not exhibit any
experimental evidence of a pronounced atomic rearrangement.
As in the case of the (100) surface of the same systems the interpretation
of such a vibrational behaviour in terms of Fermi surface effects
should be supported
by the presence of electronic surface states connected by nesting wave vectors
comparable with the critical ones resulting from the HAS spectra.
However, angular resolved photoemission (ARP) studies carried out by Kevan
and coworkers~\cite{JEO88,JEO89,GAY89} do not show any evidence of
parallel segments of
the Fermi-surface contours and thus a nesting origin of the
HAS anomalies was ruled out in these analyses.\\

The jigsaw puzzle was even more complicated by the data from a
very careful high
resolution electron energy loss spectroscopy (HREELS) investigation
on hydrogen covered W(110)~\cite{BAL93,BAL94}: The experimental points
pin down the dispersion of the small anomaly (in perfect agreement with
the HAS data), but not of the huge indentation.
Thus, the latter is only visible in helium scattering. Moreover,
the HREELS
spectra for the high energy adsorbate vibrations, which are not accessible to
the HAS experiment, are also characterized by an anomalous behaviour.
At a coverage of 1 ML of hydrogen the peaks
due to the
adsorbate vibrations parallel to the surface are no more resolved
in the HREELS spectra as
{\em single} inelastic structures but give rise to a broad continuum.
These features were explained by Balden {\it et al.}~\cite{BAL94}
as being due to the fact that
at that coverage the hydrogen atoms form a quantum fluid. Within
the framework of this picture
the sharp anomaly is an elementary excitation of this fluid similar to
a roton~\cite{FEY54}. We like to stress, however, that this interpretation
of the huge indentation fails because
it involves vibrations of the hydrogen, and therefore predicts strong
isotopic effects, which have not been
observed in the HAS experiments~\cite{HUL92,HUL93,LUE94}. Moreover, to
the best of our knowledge the
typical dispersion of a
roton is not as sharp as the one seen in the case of H/W(110) and H/Mo(110).
While we question the "roton idea", we feel that
the experimental evidence for a liquid phase is indeed strong and calls
for a detailed analysis.\\

In this paper we review the results of a recent {\it ab-initio}
study~\cite{KOH94a} which provide new insight into these unusual and exciting
phenomena.
Our results for the
atomic and electronic structure of the clean and H-covered Mo(110) surfaces
are presented and discussed in view of the two anomalies and of the
ARP data. Finally, in the Conclusion we
comment on some still open question.\\

\section{R\lowercase{esults}}
We employed the
density-functional theory (DFT) together with the local-density approximation
for
the exchange-correlation energy functional~\cite{CEP80}, using the
full-potential
linearized augmented plane wave (FLAPW) method~\cite{BLA85} for the
self-consistent evaluation of the Kohn-Sham equation, total energy, and
forces~\cite{YU91,KOH94}.\\

\begin{figure}
  \leavevmode
  \includegraphics{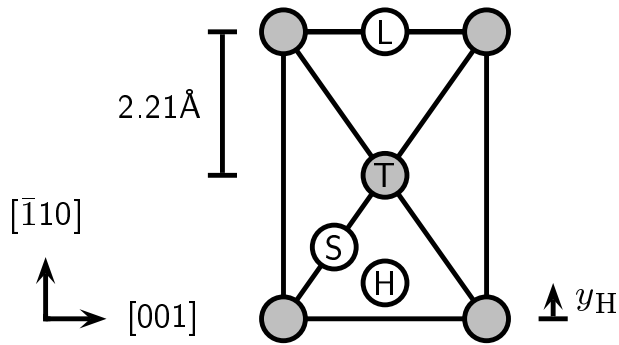}
  \vspace*{7.0cm}
\caption{Surface geometry of Mo\,(110) with the long-bridge ({\sf L}),
short-bridge ({\sf S}), hollow ({\sf H}) and on-top ({\sf T}) sites
marked. $y_{\rm H}$ is
the $[\bar{1}10]$-offset of the hollow position from the long-bridge site.}
\label{geometry}
\end{figure}
The surfaces are modeled by a seven layer slab for the
Mo(110) substrate repeated periodically with a separation of
8.8 \AA~vacuum.
A mesh of 64 equally spaced points in the surface Brillouin zone (SBZ)
is employed for the integration in the reciprocal space. The muffin-tin
radii are chosen to be 1.27 \AA~ and 0.48 \AA~for Mo and H, respectively.
The kinetic-energy cutoff for the plane wave basis needed for
the interstitial region is set to  12~Ry, and the $(l,m)$ representation
(inside the muffin tins) is taken up to $l_{\rm max}= 8$.
For the potential expansion we use a plane-wave cutoff energy
of 64~Ry~and a $(l,m)$ representation with $l_{\rm max}= 4$.
The core states as well as the valence states are treated
non-relativistically. The calculated in-plane lattice
constant is 3.13~\AA~without including zero point vibrations,
which compares well with the measured bulk lattice parameter
(3.148~\AA~at room temperature~\cite{KAT79}).
The values of the surface relaxation
for the clean surface (see Table~\ref{tab:relax}, first column)
are in excellent agreement  with
those calculated by Methfessel {\it et al.}~\cite{MET92} who
employed the full-potential linearized muffin-tin orbitals
approach~\cite{MET89}.\\

Our first goal is the determination of the geometry of the system, of
the relaxation parameters, and of
the energetics of the possible hydrogen adsorption sites
for the full monolayer coverage. For this
coverage there is one hydrogen atom per unit cell and the four possibly
important
adsorption positions are depicted in Fig.~\ref{geometry}.
They are the long-bridge ({\sf L}), short-bridge ({\sf S}),
hollow ({\sf H}) and on-top ({\sf T}) sites.
\begin{table}
\begin{tabular}{c|c|cccc}
& clean & hollow & long bridge & short bridge & on top \\
\hline
$d_{\rm H-Mo}$ (\AA)        & ---    &$  1.07$&$  1.08$&$  1.32$&$  1.76$\\
$\Delta d_{12}~(\%d_{0})$   &$ -4.5 $&$ -2.1 $&$ -1.9 $&$ -2.8 $&$ +1.8 $\\
$\Delta d_{23}~(\%d_{0})$   &$ +0.5 $&$ +0.1 $&$ -0.1 $&$ +0.3 $&$ -1.2 $\\
$\Delta d_{34}~(\%d_{0})$   &$  0.0 $&$ -0.1 $&$ -0.3 $&$ -0.3 $&$ +0.1 $\\
$\Delta E_{\rm ad}$ (eV)    & ---    &$  0.0 $&$ -0.23$&$ -0.28$&$ -1.11$\\
\end{tabular}
\caption{Calculated geometries and adsorption-energy differences  for clean
Mo\,(110) and for the H-covered surface. For the latter the results for
the long-bridge, short-bridge, hollow, and on-top sites (see Fig. 1) are
compiled. The height of the hydrogen atom above
the surface is denoted as $d_{\rm H-Mo}$ and $\Delta d_{ij}$ is the percentage
change of the
inter-layer distance between the $i$-th and the $j$-th layer with
respect to the bulk inter-layer spacing $d_{0}$.}
\label{tab:relax}
\end{table}
The results of the calculations are collected in
Table~\ref{tab:relax}.
Clearly, the hollow site appears as the energetically most favorable
site, while
the difference in energies between the long-bridge and short-bridge positions
is small. On the other hand,
a hydrogen in the on-top location appears definitely unstable.
To find the exact position of the adsorbate we start with the hydrogen
in the long-bridge position and we allow its relaxation according
to the forces.
The hydrogen relaxes directly into the hollow site finding its stable
geometry at a position $y_{\rm H} = 0.55 \pm 0.01$\AA~(see
Fig.~\ref{geometry}).
We do not find a pronounced top-layer-shift
reconstruction is which is in agreement with an analysis of low-energy electron
diffraction experiments \cite{ALT87}. A comparison with the calculated
clean surface geometry shows that adsorption of hydrogen produces a
reduction of the inward relaxation as expected. In the case
of the on-top
adsorption site the relaxation becomes even outwards.
Note that the numerical error for all calculated
inter-layer spacings is about $\pm 0.3~\%d_0$.
Table~\ref{tab:relax} shows that the
differences in energy among the adsorption sites are similar to
those found for other systems, H/Pd(100) and H/Rh(100)~\cite{WILxx,FEIxx}.
We should point out that the investigation of Balden
{\it et al.} was carried out on H/W(110) and HREELS experiments on H/Mo(110)
are in progress. However, our preliminary results on H/W(110) show
similar adsorption energies for the four sites.
Therefore, our results do not seem to rule out the
presence of a liquid-like phase
of the hydrogen
atoms on W(110).\\

\begin{figure}
    \leavevmode
    \vspace*{8cm}
    \includegraphics{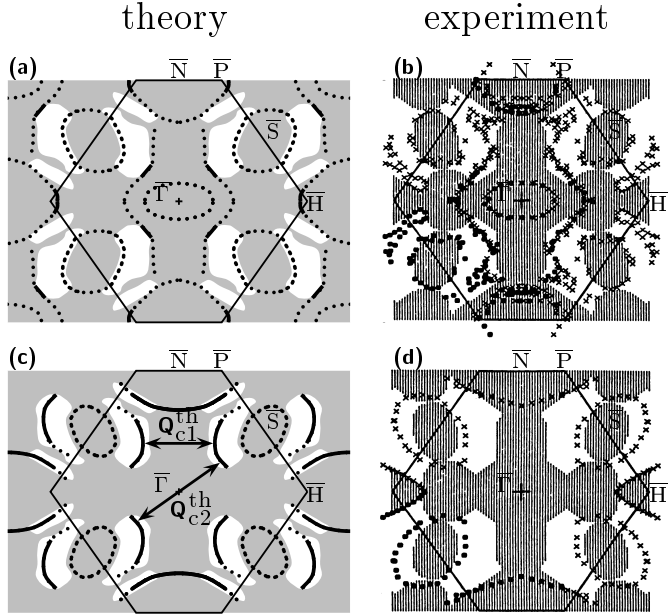}
\caption{Fermi surfaces of clean Mo\,(110) [upper panels]
and Mo\,(110) with a full
monolayer of hydrogen [lower panels]. The experimental results are from
Ref.~\protect\citenum{JEO89}.
Shaded areas are the projection of the bulk Fermi
surfaces
onto the (110) surface Brillouin zone. The solid (dotted) lines of the
theoretical results denote
surface resonances or surface states which are localized by more than
60 \% (30 \%) in the two top Mo layers.
${\bf Q}_{\rm c1}^{\rm th}$ and
${\bf Q}_{\rm c2}^{\rm th}$ are the critical wave vectors, predicted by the
calculations.}
\label{disper}
\end{figure}

After having determined the energetically most stable geometry we
calculated the electronic properties of the surface in both phases,
clean and covered with a full monolayer of hydrogen. The theoretical
results for the
Fermi-surface contours are
collected in Fig.~\ref{disper} (left panels) together with experimental
ARP data of Jeong
{\em et al.}~\cite{JEO89} (Fig.~~\ref{disper}, right panels).
The solid (dotted) lines
label surface resonances or surface states which are localized by more
then 60 \% (30 \%) in the two top Mo layers.
For the adsorbate system we present the results for the hollow position.

The shaded area is the projection of the
bulk Fermi surface onto the SBZ of the (110) surface. For the experimental
analysis Jeong {\em et al.} calculated this projection using a
semiempirical tight-binding interpolation scheme~\cite{PAPA86}. The
differences between the results (left and right panels in Fig.~\ref{disper})
for the projected bulk Fermi surface (shaded areas) may arise from the
lack of self-consistency of the tight-binding approach. We notice
that in Fig. 8 of Ref.~\citenum{JEO89} Jong {\em et al.} presented
results with their Fermi level shifted up by 200 meV. In this case
a good agreement of their results with our DFT-LDA is achieved.
For the clean surface (see Fig.~\ref{disper}a) our calculations identify
four bands, which are highly localized at the surface. They give rise to
Fermi-surface contours of which
one is centered at $\overline{\Gamma}$, one at
$\overline{\rm S}$, one at $\overline{{\rm N}}$, and one which runs
more or less parallel to the $\overline{\Gamma{\rm N}}$ direction.
The latter
has ($d_{3z^{2}-r^{2}}, d_{xz}$) character, and also the other
surface features originate from Mo $d$ bands.
All of these bands have been detected experimentally, and for all of
them the calculated and measured ${\bf k}$-dependence is in very good
agreement (compare Figs.~\ref{disper}a and b). On the other hand, the
comparison
for the adsorbate systems (see Figs.~\ref{disper}c and d) shows
a clearly worse agreement.

\begin{figure}
    \leavevmode
    \vspace*{9cm}
    \includegraphics{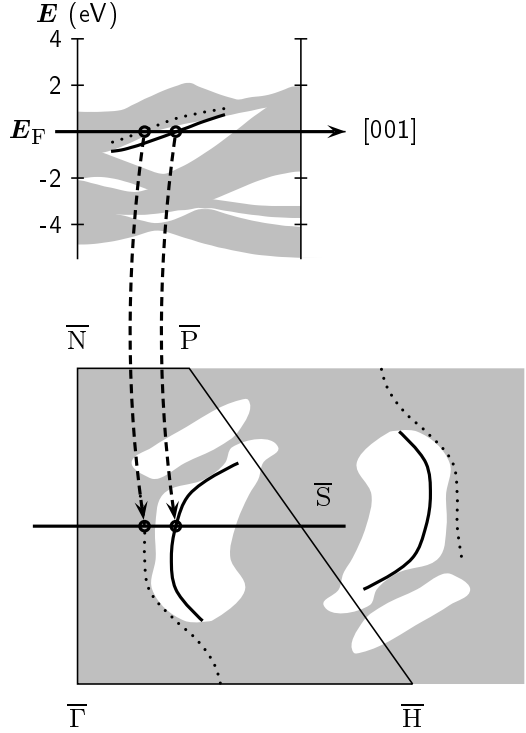}
\caption{Band structure (upper figure) and Fermi surface (lower figure)
of clean Mo(110) (dotted curves) and
H/Mo(110) (solid curves). The shaded areas are the projection of the
Fermi surface onto the SBZ of the bcc (110) surface.}
\label{bandef}
\end{figure}

At first we discuss the changes of the theoretical Fermi surface induced
by the hydrogen adsorption (Fig.~\ref{disper}a and c), and then we return
to the experimental results.
For the present discussion the most important effect
is that the ($d_{3z^{2}-r^{2}}, d_{xz}$) band is shifted away from the
$\overline{\Gamma{\rm N}}$ line. This effect can be better viewed in
Fig.~\ref{bandef}.
On the clean surface the ($d_{3z^{2}-r^{2}}, d_{xz}$) (dotted line) state is
present,
but it is buried in the bulk band density of states. After adsorption
of a full monolayer of hydrogen, a significant fraction of the
band has been shifted into the stomach shaped band gap. This shift reflects
the change in the surface potential due to the
screening of the hydrogen adatom induced perturbation
by the easily polarizable surface $d$ electrons.
In the {\em frontier orbital} picture proposed by Hoffmann~\cite{HOF88} the
hydrogen $s$-orbital interacts mainly with  occupied states close
to the bottom of the $d$-band. Whereas the corresponding bonding states are
situated 5 eV below the Fermi level, the antibonding states, which
would be partly occupied, are at at higher energies (about
4~eV above the Fermi energy) and they will dump their electrons at the Fermi
level. As a consequence, the
corresponding occupation redistribution to states at
$E_{\rm F}$ gives raise to the observed decrease of the surface
potential a shift of the entire ($d_{3z^{2}-r^{2}}, d_{xz}$) band
to lower energies occurs.
As this band disperses upward when
${\bf k}$ increases away from
$\overline{\Gamma}$, the Fermi energy cuts the shifted band at wave vector
with a larger component along $\overline{\Gamma {\rm H}}$. In addition, we
notice the important fact that no strong changes
in the form of the Fermi-surface
orbits are observed in our results when we (artificially) place the hydrogen
atom
in the short-bridge or long-bridge positions.

Figures~\ref{disper}c and~\ref{bandef} reveal that the Fermi-surface
contours associated to the ($d_{3z^{2}-r^{2}},d_{xz}$)-like band
run parallel to the $\overline{\Gamma {\rm N}}$ direction
and perpendicular to $\overline{\Gamma {\rm S}}$ in significant
parts of the SBZ giving rise to a quasi-one-dimensional nesting.
The {\bf k}-vectors connecting these bands in different
sections of the Brillouin zone are labeled in Fig.~\ref{disper}c
with ${\bf Q}_{\rm c1}^{\rm th}$ and ${\bf Q}_{\rm c2}^{\rm th}$.
We recall that in DFT the highest occupied
Kohn-Sham level of the self-consistent
$N$-electron calculation corresponds to the electron
chemical potential, i.e., the Fermi {\em level}.
However, the Kohn-Sham Fermi {\em surface} is (in principle) not an
observable. On the other hand, the agreement between
Fig.~\ref{disper}a and b and the fact that the
wave function character does not change significantly upon hydrogen
adsorption support the hypothesis
that the difference between the Kohn-Sham Fermi surface and the
true Born-Oppenheimer Fermi surface is not very important for this system.
We therefore conclude that the nesting instability predicted
by the Fermi surface of Fig.~\ref{disper}c can be trusted. Moreover,
the dispersion of the ($d_{3z^{2}-r^{2}}, d_{xz}$) band in
Fig.~\ref{bandef} is flat and the associated electronic density
of states at the Fermi level is high. Therefore, a large number of
electrons are involved in the nesting and the mechanism
increases its efficiency. Because of this aspect and due to
its nearly one-dimensional nature we expect important
consequences~\cite{TOS75}.
Such an instability may produce a static rearrangement of the atoms
provided that the electronic energy gain is stronger than the elastic
energy cost needed to distort the atomic lattice. This is the case
of the (100) surface of Mo and W, where this kind of distortion
accompanied by the appearance of a charge density wave occurs
below a critical temperature. Because the
(110)\,surface is the closest packed and most stable bcc surface
such a Peierls distortion is hampered and somewhat unlikely.
When the system does not undergo a static distortion,
strong dynamical consequences (similar to a dynamical
Jahn-Teller effect in molecules) will take place due to thermally excited
electron-hole pairs with a wave vector equal to the critical wave vectors
and their coupling to surface phonons. In other words, the electronic
system is close to an instability supported by a surface phonon,
but the instability can not be removed via a Peierls distortion because
of the large cost in elastic energy. Therefore, we expect low frequency
oscillations of the electronic charge density near the Fermi level.\\

Because of these theoretical evidences we suggest the following
interpretation of the experimental features.
\begin{figure}
    \leavevmode
    \vspace*{9cm}
    \includegraphics{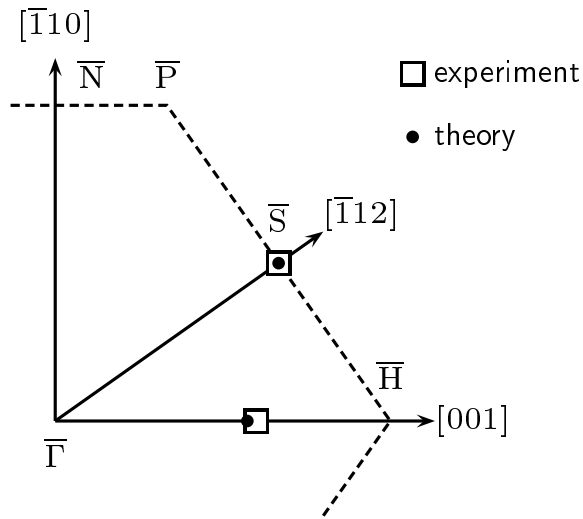}
\caption{{\bf k}-points loci of the HAS-measured (open squares) and calculated
(full dots)
critical wave vectors in the SBZ for H/Mo(110).}
\label{locusan}
\end{figure}
It is well know that
for metal surfaces the turning point of the helium atoms in a HAS
experiment is at a
distance of about 3-4~\AA~in front of the surface and that
the scattering potential depends on the electron density. The probe
particles thus will sense electron
charge density oscillations close to the Fermi level~\cite{KADxx}.
For the system Mo\,(110) at the critical wave vectors these oscillations
are associated with electron-hole pair excitations involving the
($d_{3z^{2}-r^{2}}, d_{xz}$) band.
We therefore propose that the two bands seen by HAS are due to a
hybridization between lattice vibrations and electron-hole pairs.
Thus these studies represent an experimental manifestation of
a nearly one-dimensional Kohn anomaly.
While the small dip has
a stronger phonon like character, the sharp and
giant anomaly is due to a more electron-hole like excitation.
Balden~\cite{BALBH} has recently found
that the small indentation has indeed a temperature dependence like
that of a phonon:
Its frequency increases with increasing temperature.
Furthermore, it has been observed~\cite{HUL93,LUE94} that the anomalies
persist in the
phonon dispersion curves  along directions deviating from
$\overline{\Gamma {\rm H}}$, which is well
understandable in terms of the theoretical results of Fig.~\ref{disper}c,
because nesting is present also for the directions in the surface Brillouin
zone out of the [001] azimuth.
Because of the form of
the ($d_{3z^{2}-r^{2}}, d_{xz}$) band Fermi-surface contour, the anomalies
should be particularly strong along $\overline{\Gamma {\rm S}}$ in agreement
with
the available experimental data~\cite{LUE94}. In Fig.~\ref{locusan} we
show the comparison between the experimental points in the SBZ,
where the anomalies have been measured, and the points, where large
section of the calculated Fermi surface contours are nearly parallel.
The agreement is very good. We expect a weakening of the
indentations for directions between $\overline{\Gamma {\rm H}}$ and
$\overline{\Gamma {\rm S}}$ because of the less effective nesting.
Unfortunately, no dispersion curves for the other two wave vectors mentioned in
the literature~\cite{HUL93,LUE94} are available.
A careful reanalysis of the experimental data is in progress~\cite{HUL94}.

Our interpretation of the HAS results also explains why in HREELS only the
Rayleigh mode with the small dip can be detected. In the impact regime
electrons
are scattered mainly by the ion cores and map the phonon dispersion curves.
However, they are practically insensitive to
electronic charge density oscillations~\cite{KADxx}. Because
the sharp and giant
anomaly has only little vibrational character, it is practically impossible
to excite it by high-energy electrons.

An open question is why the ARP measurements  for the hydrogen
covered Mo\,(110) and the theoretical Kohn-Sham Fermi surface are so much
different. The results of Fig.~\ref{disper}c are due to photoelectrons
which carry the information about the difference of the $N$ particle ground
state and the $N-1$ particle system with a hole at $|\epsilon_{F}, {\bf
k}\rangle$.  Typically an analysis surmises that the electronic
system is in its ground state and that vibrational excitations
as well as interactions between electronic and vibrational
excitations can be neglected.
The theoretical results shown in Fig.~\ref{disper}c, which are obtained
under exactly this hypothesis, disclose that for this system the assumption
may not be valid. In fact, the calculations predict
highly localized parallel bands at the Fermi surface with a
one-dimensional nesting vector. As a consequence,
at ${\bf Q}_{\rm c1}$ and ${\bf Q}_{\rm c2}$ many electron-hole pairs
will be thermally excited, and these will couple to phonons.
This produces a breakdown of the Born-Oppenheimer approximation.
While the calculations identify a strong nesting,
the ARP experiments investigate a system which has already reacted
to this nesting instability.
When a static distortion is frustrated,
ARP will measure a state with a significant number of low energy excitations
(see Fig.~\ref{bandef}),
and in particular at the critical wave vectors the electron and lattice
dynamics
cannot be decoupled.
In fact, we note that the experimental Fermi surface
of the H covered surface seems to display a lack of the translational
symmetry associated with the $(1 \times 1)$ SBZ (see the band circuits at the
$\overline{\rm S}$ point in Fig.~\ref{disper}d). We cannot rule out, however,
that this is due to inaccuracies in the experimental
data or analysis. We are not aware that for any other surface such a serious
breakdown of the
Born-Oppenheimer approximation (compare Figs.~\ref{disper}c and d)  has
been observed so far. Obviously, the static distortion of
the W\,(100) surface is also induced by a Kohn anomaly, but after the
reconstruction has taken place the Born-Oppenheimer approximation is
valid again. Some doubts remain at this point if an alternative  explanation
exists reconciling the differences of the experimental
ARP data and the calculated results.

It is interesting to note that the energy barrier between two hollow sites is
about 0.2 - 0.3 eV. This is, however, a rough estimate, because the
results of Table~\ref{tab:relax} do not correspond to a single H-atom hopping
event but to a displacement of the whole H layer. Nevertheless, it is in fact
likely that at room temperature the full hydrogen monolayer might be
dynamically strongly disordered: At a snapshot picture some H atoms may
occupy the tip down triangles and others the tip up triangles.
In this way a liquid like behaviour may result
We note, however, that
because of the lack of isotopic effects in
the characteristics of the phonon branch indentations,
we discard any direct involvement of hydrogen vibrations
in the anomaly mechanism.

\section{C\lowercase{onclusion}}
In conclusion, the calculated Kohn-Sham Fermi surface can explain the
HAS and HREELS experiments. It is interesting that the same
band which is responsible for the pronounced anomaly identified above
for  Mo\,(110) produces in fact notable features
of all group VIA materials. In particular we mention the spin-density wave
in bulk Cr and the phonon anomalies along the [001] direction in bulk W and Mo.
Also the reconstruction
of Mo\,(100) and W\,(100) is induced by the nesting of this band.
What makes the anomaly at the (110) so particularly strong is that the
adsorption of hydrogen shifts it into a region of ${\bf k}$-space where
it becomes localized at the surface. Thus the band becomes two- and the
nesting one-dimensional.
Neither the hydrogen wave functions nor the hydrogen vibration are
{\em directly} involved in the resulting anomalies. The effect of hydrogen is
mainly to change the potential at the surface and to shift the Mo surface
states. Thus, it is well possible, that other adsorbates can influence
in a similar way the surface vibrations and electron-phonon coupling.
We also expect that  for the clean surface a  weak
signature of the ($d_{3z^{2}-r^{2}}, d_{xz}$) band should exist at ${\bf Q}
\approx
0.6$~\AA$^{-1}$. However, for the clean surface the band
is not a surface state but a broad and less localized surface resonance.
The thermal excitations of
electron-hole pairs at H/Mo\,(110) and the  suggested breakdown of the
Born-Oppenheimer approximation
call for additional photoemission experiments, in which the Fermi surface
is carefully studied also at lower  temperatures.
Furthermore, calculations of the electron-phonon coupling, electron-hole
excitations and frozen phonons at these surfaces would be very
helpful.

\section{A\lowercase{cknowledgments}}
We thank M. Balden, J. L\"udecke, E. Hulpke, and E. Tosatti for
stimulating discussions.

\end{document}